\documentclass[12pt]{article}
\usepackage{graphicx}
\usepackage{epstopdf}
\usepackage{graphicx}
\usepackage{amsmath}
\usepackage{bm}

\def\itmb{\begin{itemize}}
\def\itme{\end{itemize}}
\def\enmb{\begin{enumerate}}
\def\enme{\end{enumerate}}
\def\eqnb{\begin{equation}}
\def\eqne{\end{equation}}

\title{On the Amplitude of External Perurbation and \\
Chaos via Devil's Staircase\\
in Muthuswamy-Chua System}

\author{Sadataka Furui  \\
 Graduate School of Teikyo University\\
2-17-12 Toyosatodai, Utsunomiya, 320-0003 Japan 
{\thanks {\textit{E-mail address:} furui@umb.teikyo-u.ac.jp }}
\and
Tomoyuki Takano\\
IIX inc, Location 7F Daiichi Life Ins. Bld. Annex, \\
2-7-12 Nishi-Gotanda, Shinagawa-ku, Tokyo 141-0031, Japan
{\thanks {\textit{E-mail address:} takano\_t@iix.co.jp}}
}

\begin{document}
\maketitle


\begin{abstract}
We recently analyzed the voltage of the memristic circuit proposed by Muthuswamy and Chua by adding an external sinusoidal oscillation $\gamma\omega \cos\omega t$ to the ${\dot y}(t)\simeq {\dot i_L}(t)$, when the ${\dot x}(t)\simeq {\dot v_C}(t)$ is given by $y(t)/C$.

When $f_s<f_d$ we have observed  that the H\"older exponent of the system with $C=1$ is larger than 1, and that of the system with $C=1.2$ is less than 1. The latter system is unstable, and the route to chaos via the devil's staircase is observed. 

Above the mode of $f_d=1, f_s=1$ observed at $\omega\simeq 0.5$, we observed a mode of $f_d=1, f_s=2$ at $\omega\simeq 1.15$ and $\simeq 1.05$, in the case of $C=1$ and 1.2, respectively,
and a mode of $f_d=2, f_s=3$ at $\omega\simeq 0.85$ and $\simeq 0.78$, in the case of $C=1$ and 1.2, respectively.
At high frequency of $f_s$, there is no qualitative difference in the stability of the oscillation for 
$C=1$ and $C=1.2$

\end{abstract}


\section{Introduction}
\noindent In 1998, Dos Santos studied the system that follows
\[
\Phi_{n+1}=\Phi_n+2\pi \Omega+c \sin\Phi_n, 
\]
which is obtained from Van der Pol's oscillator perturbed by external
oscillation\cite{Santos98}.  He defined $f(\Phi)=\Phi+2\pi\Omega+c\sin\Phi$ on $[0,2\pi]$, which is known as the
Morse-Smale diffeomorphic function when $c<1$ \cite{Devaney89}.
When we calculate the derivative with respect to $\Phi$, we obtain
\begin{equation} \label{phieq}
f'(\Phi)=1+c\cos\Phi.
\end{equation}
When $c>1$, there appear several values of $\Phi$, where $f'(\Phi)=0$. When there are degenerate
states in nonlinear systems, transition to another manifold of orbits becomes possible and complicated chaotic behavior becomes observable. The transition to chaos occurs via appearance of devil's staircase pattern in the 
structure of frequency of oscillation pattern.

 In non linear circuit, mechanisms of appearance of oscillation including devil's staircase pattern of oscillation were studied\cite{CYY86}.
In 2010, Muthuswamy and  Chua\cite{MuCh10} showed that a system with an inductor, 
capacitor and non-linear memristor can produce a chaotic circuit. The three-element Muthuswamy-Chua
system with the voltage across the capacitor $x(t)=v_C(t)$, the current through the inductor
 $y(t)=i_L(t)$ and the internal state of the memristor $z(t)$\cite{ChKa76}, satisfy the equation
\begin{eqnarray}  
\dot x&=&\frac{y}{C},\nonumber\\
\dot y&=&\frac{-1}{L}[x+\beta(z^2-1)y],\\
\dot z&=&-y-\alpha z+y\,z. \nonumber
\end{eqnarray}
\label{memristor}

The flow curvature manifold of the memristor was studied in \cite{GLC10} and \cite{LlVa12}.
They modified the sign of $\dot z$ from that of \cite{MuCh10}, which does not make differences in the topological structure, and adopted the system as follows;
\begin{eqnarray}
 \dot x&=&y,\nonumber\\
 \dot y&=&\frac{-1}{3}[x+\frac{3}{2}(z^2-1)y],\\
 \dot z&=&y+\alpha z-y\,z.\nonumber
\end{eqnarray}

They defined the vector field $X$ 
\[
X=y\frac{\partial}{\partial x}+\frac{-1}{3}[x+\frac{3}{2}(z^2-1)y]\frac{\partial}{\partial y}+(y-\alpha z-y\,z)\frac{\partial}{\partial z}
\]
and observed that when $\alpha=0$, $H=x+log(1-z)$ is a first integral, and the system has the Darboux type integrability.

When $\alpha\ne 0$, they did not find Dorbeaux type first integrals.
Zhang and Zhang\cite{ZhZh13} showed, if $\alpha>0$ and the system has a periodic orbit or a chaotic attractor, the orbit must intersect both the planes $z=0$ and $z=-1$ infinitely many times as tends to infinity. As a byproduct, they got unstable invariant behaviors under small perturbations.

We added a small perturbation of $F=\gamma \sin\omega t$ to the Muthuswamy-Chua system \cite{FuTa13} and considered the coupled differential equation, using parameters $L=3.3, \alpha=0.2, \beta=0.5$ and $C=1$ and $C= 1.2$.
\begin{eqnarray}  
\dot x&=&\frac{y}{C},\nonumber\\
\dot y&=&\frac{-1}{L}[x+\beta(z^2-1)y+\gamma \sin\omega t],\\
\dot z&=&-y-\alpha z+y\,z. \nonumber
\end{eqnarray}
\label{perturbedmemristor}

When the stability condition $\displaystyle \frac{d^2y}{dt^2}=0$ is chosen, 
\begin{equation}
\frac{-1}{L}(\frac{y}{C}+\beta(z^2-1)\frac{dy}{dt}+2\beta \frac{dz}{dt} z\,y+\gamma\omega\cos\omega t)=0
\end{equation}
is obtained.
The changing of the surface of the solution occurs when $\displaystyle \frac{dy}{dt}=\frac{dz}{dt}=0$, or
$y(t)+C\gamma\omega\cos\omega t=0$ which is structurally same as eq.(\ref{phieq}).  We fixed $\gamma=0.2$, and chose $C=1$ and $C=1.2$.
We observed standard devil's staircase structure when $C=1$, but complicated chaotic behavior when $C=1.2$.
The qualitative behavior of the chaos is a function of the amplitude of the perturbation in Dos Santos's system  
and that of perturbed memristic current are the same, but in Dos Santos's case, perturbation is $c\sin \Phi$
while in the memristic current case, it is $\gamma\sin\omega t$, in which $\omega$ is given.

\section{The single period output of the memristor}
In this section, we show the single period output $y(t)$ whose angle frequency $\omega$ is chosen in the middle of
the largest window in the bifurcation diagram  $\omega=0.59$rad/s in the case of $C=1$ and $\omega=0.535$rad/s in the case of $C=1.2$.

\begin{figure}[htb]
\begin{minipage}[htb]{0.47\linewidth}
\begin{center}
\includegraphics[width=5cm,angle=0,clip]{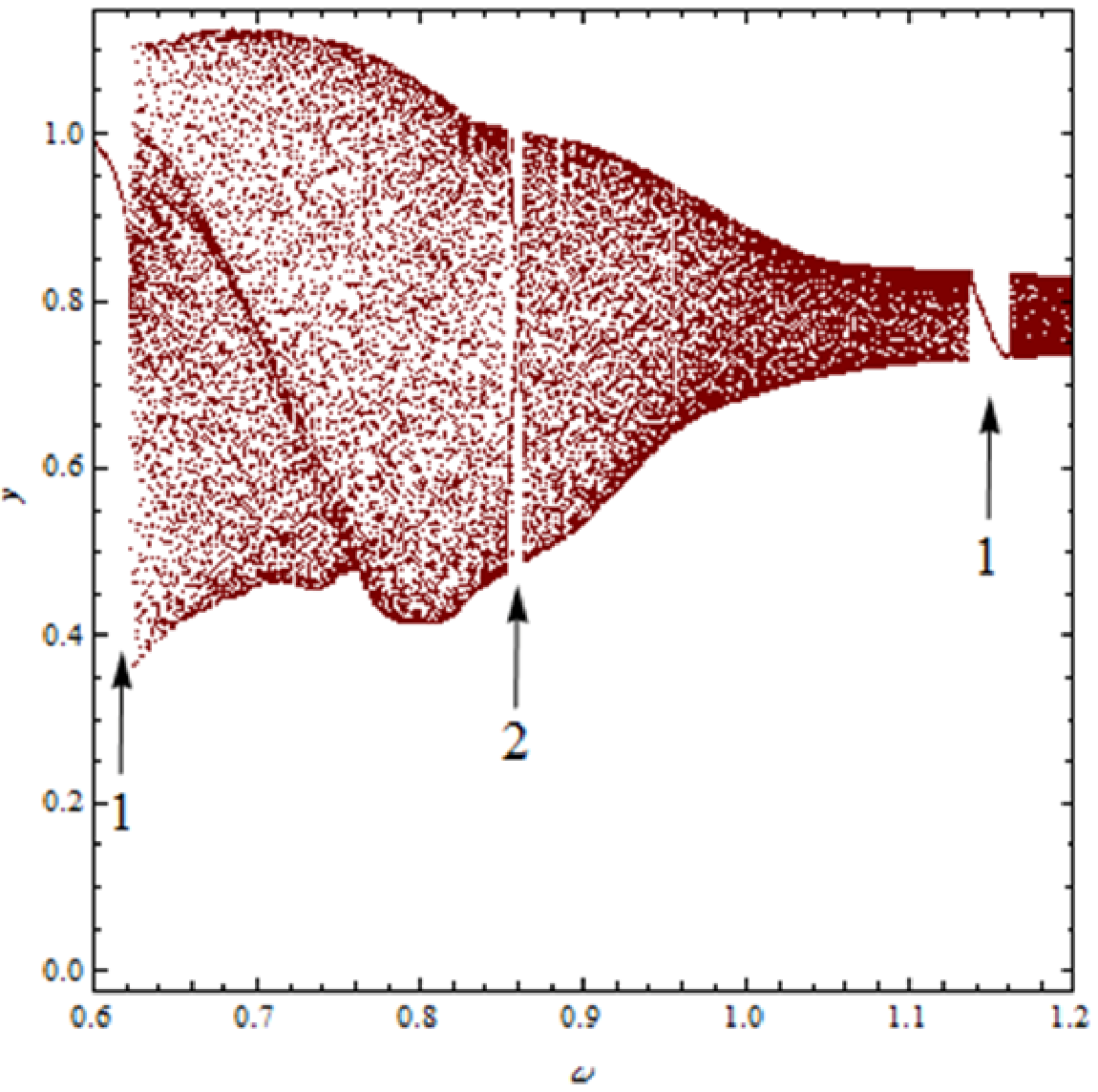}
\caption{The bifurcation diagram of the set $C=1$, $0.6<\omega<1.2$}
\label{yC1}
\end{center}
\end{minipage}
\hfill
\begin{minipage}[htb]{0.47\linewidth}
\begin{center}
\includegraphics[width=5cm,angle=0,clip]{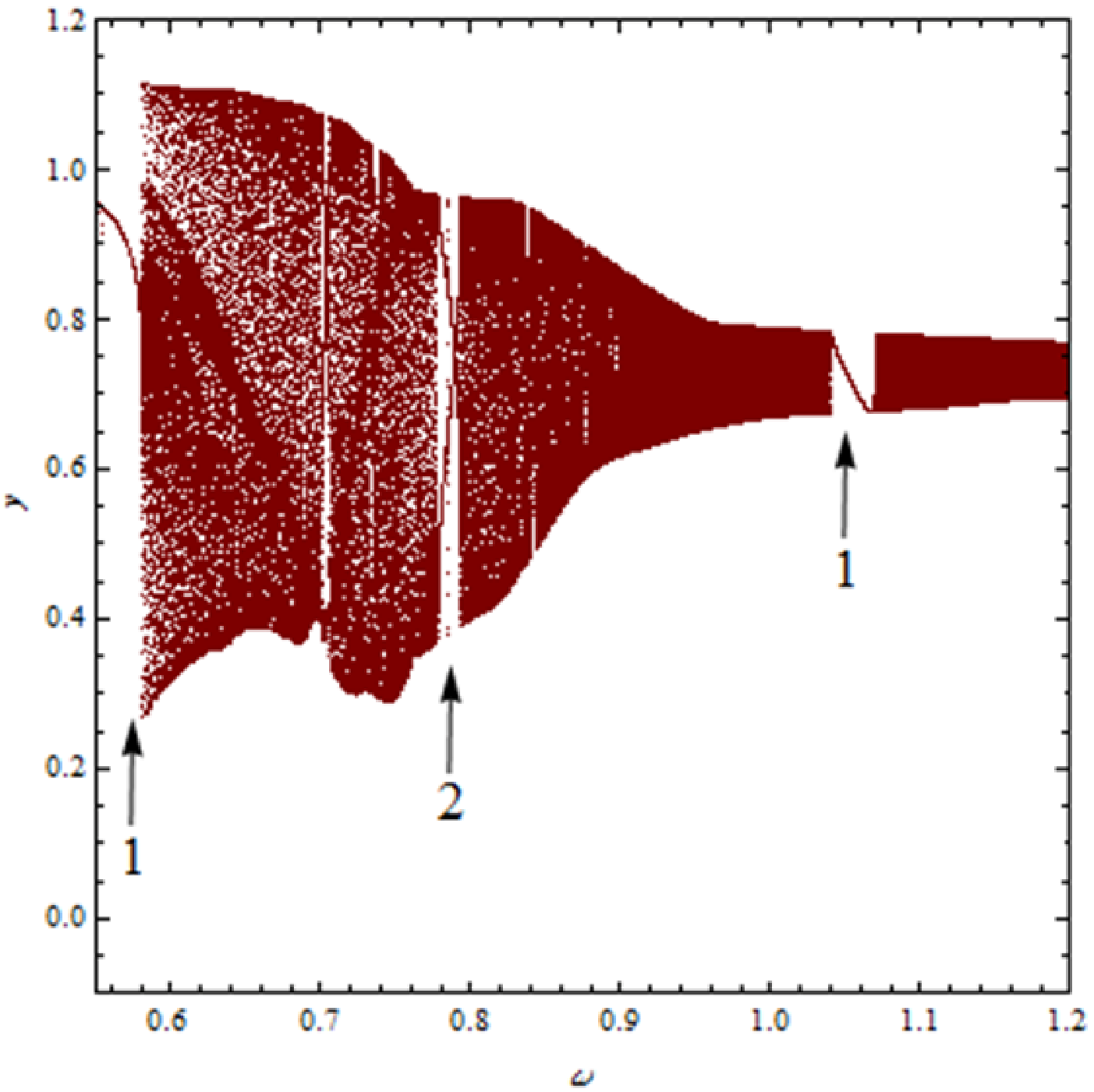}
\caption{The bifurcation diagram of the set $C=1.2$, $0.55<\omega<1.2$}
\label{yC12}
\end{center}
\end{minipage}
\end{figure}

As shown in Fig.\ref{yC1} and Fig.\ref{yC12}, above the largest window, a bifurcation diagram shows another single period output $y(t)$ and between the two single period output, there is a double period output.

\begin{figure}[htb]
\begin{minipage}[htb]{0.47\linewidth}
\begin{center}
\includegraphics[width=6cm,angle=0,clip]{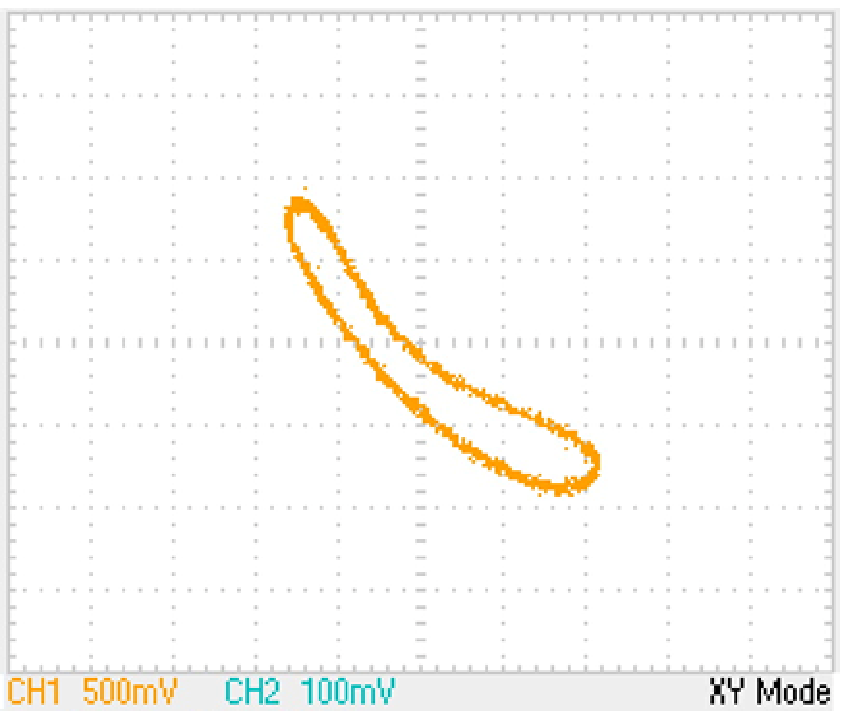}
\caption{The input w.f. $F(t)$ v.s. output w.f. $y(t)$. $f=9.35$kHz. $C=1$.}
\label{yFV1}
\end{center}
\end{minipage}
\hfill
\begin{minipage}[htb]{0.47\linewidth}
\begin{center}
\includegraphics[width=6cm,angle=0,clip]{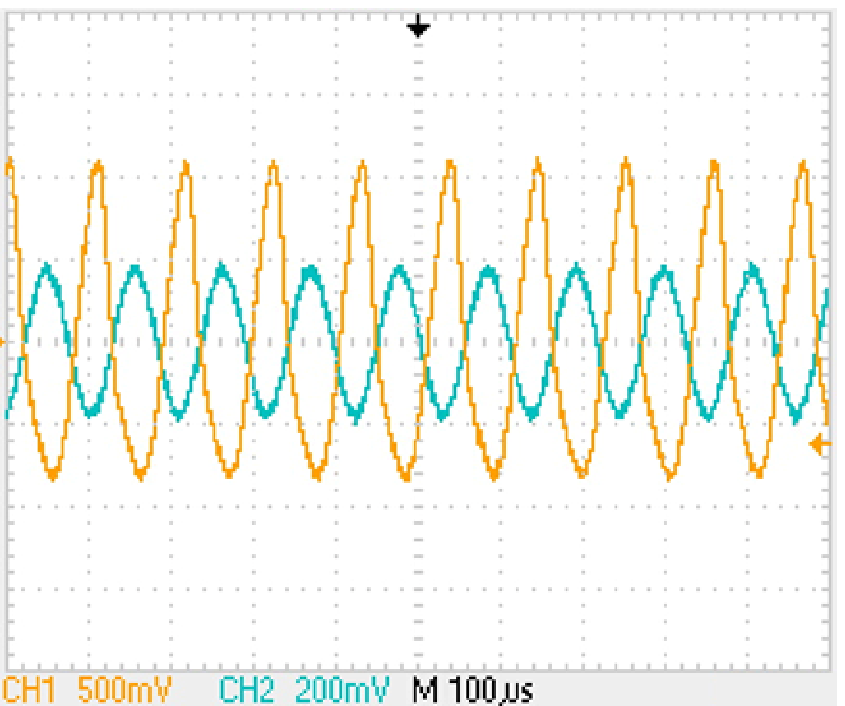}
\caption{The output w.f. $y(t)$ and input w.f. $F(t)$ (green). $f=9.35$kHz. $C=1$.}
\label{yFt1}
\end{center}
\end{minipage}
\begin{center}
\includegraphics[width=12cm,angle=0,clip]{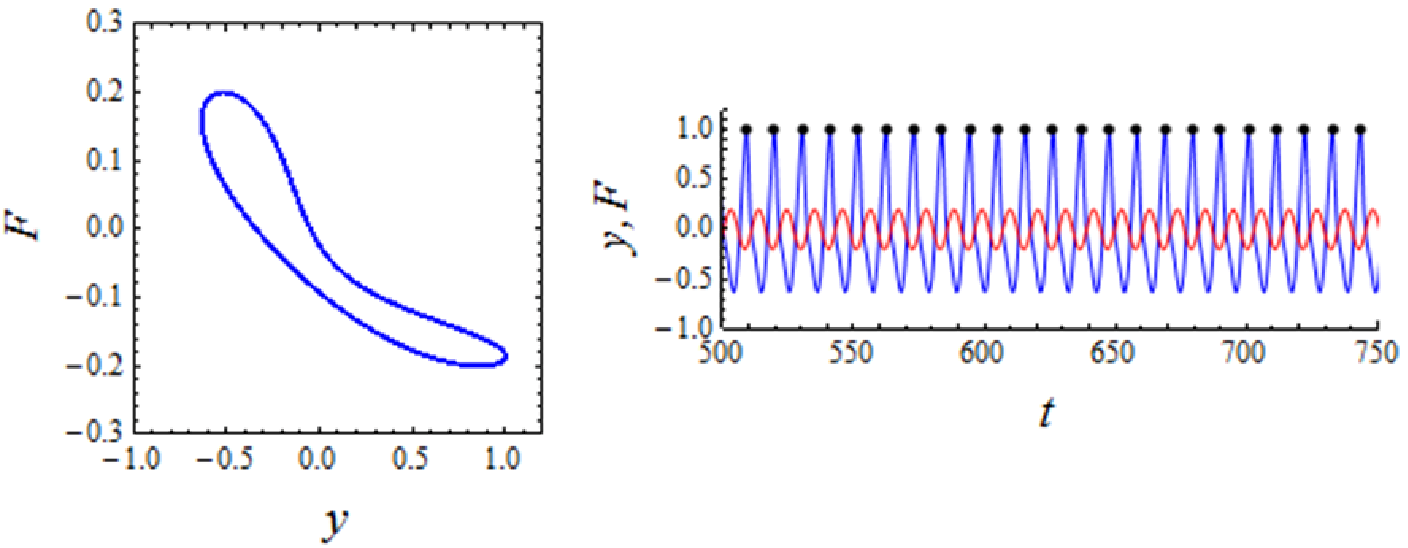}
\caption{The output w.f. $y(t)$ and the input w.f. $F(t)$ (red), $C=1$.}
\label{yFsim1}
\end{center}
\end{figure}

In Fig.\ref{yFV1} we show the experimental result of the output wave function (w.f.) $y(t)$ as a function of the input w.f.  $F(t)=\gamma\sin\omega t$.  We use $C_n=1$nF,$L_n=330$mH, $\alpha_{10kpot}=5{\rm k}\Omega$, $\beta_{5{\rm k}pot}=0.5{\rm k}$,$\gamma=200$mV in the experiment\cite{FuTa13}, anduse the notation  $C=1,L=3.3, \alpha=0.2,\beta=0.5,\gamma=0.2$ in the simulation.
The frequency in the experiment is $\displaystyle f=\frac{\omega}{2\pi}$.

In Fig.\ref{yFt1} we show the time series of $y(t)$ and $F(t)$. The computer simulations of the two w.f. are shown in Fig.\ref{yFsim1}.

\begin{figure}[htb]
\begin{minipage}[htb]{0.47\linewidth}
\begin{center}
\includegraphics[width=6cm,angle=0,clip]{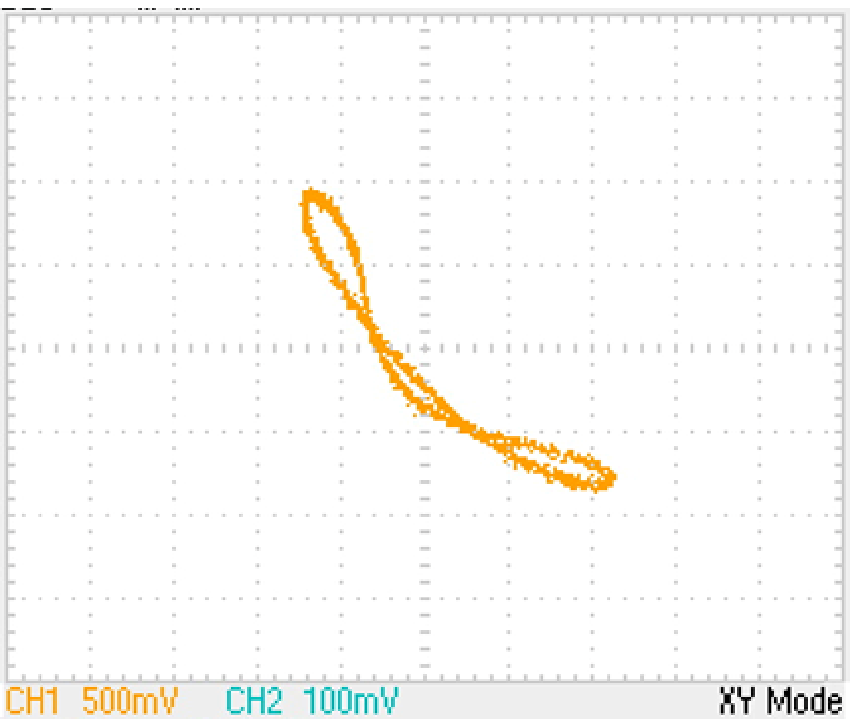}
\caption{The output w.f. $y(t)$v.s. input w.f. $F(t)$. $f=8.5$kHz. $C=1.2$.}
\label{yFV2}
\end{center}
\end{minipage}
\hfill
\begin{minipage}[htb]{0.47\linewidth}
\begin{center}
\includegraphics[width=6cm,angle=0,clip]{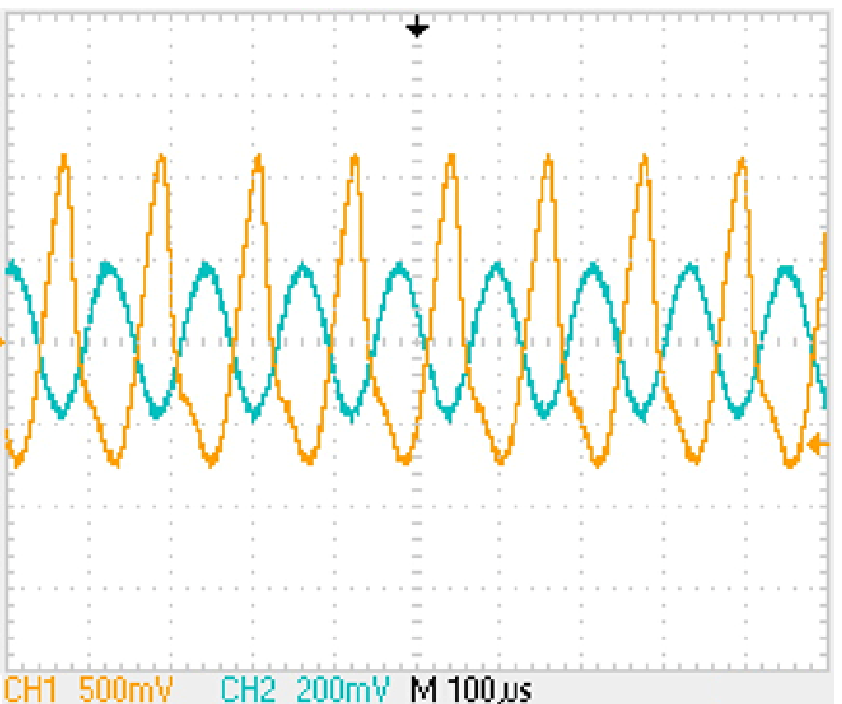}
\caption{The output w.f. $y(t)$ and input w.f. $F(t)$ (green). $f=8.5$kHz. $C=1.2$}
\label{yFt2}
\end{center}
\end{minipage}
\begin{center}
\includegraphics[width=12cm,angle=0,clip]{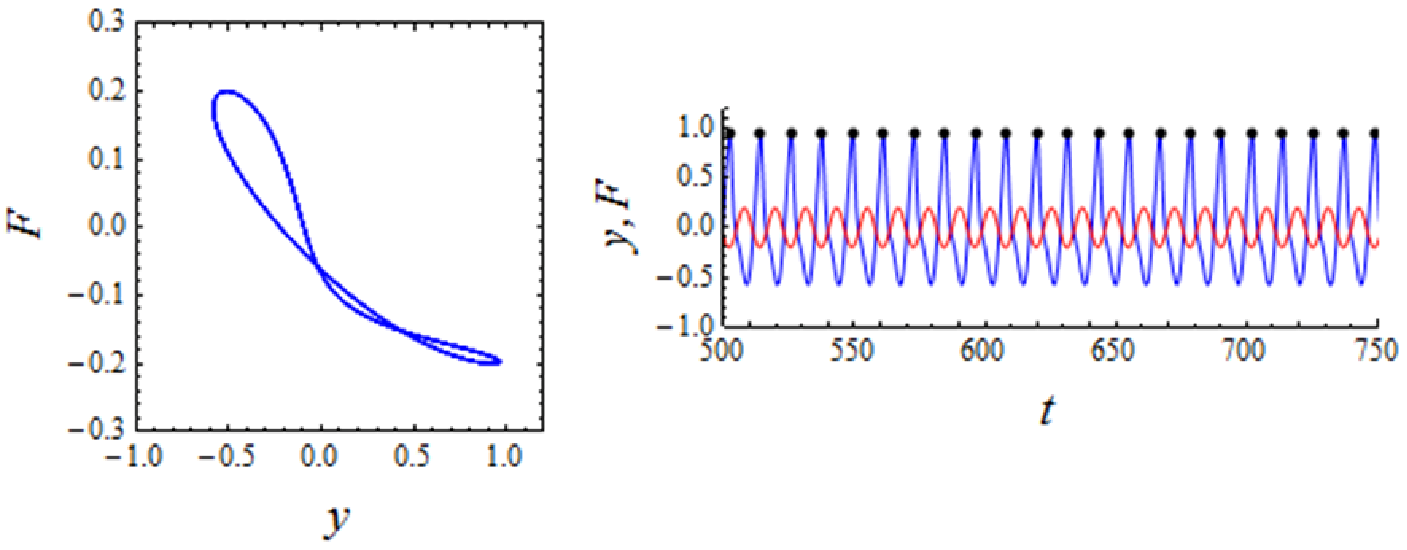}
\caption{The output w.f. $y(t)$and input w.f. $F(t)$ (red), $C=1.2$.}
\label{yFsim2}
\end{center}
\end{figure}

The corresponding data obtained by using $C=1.2$ are given in Figs.\ref{yFV2},\ref{yFt2}, \ref{yFsim2}. We use $C_n=1.2$nF,$L_n=330$mH, $\alpha_{10kpot}=5{\rm k}\Omega$, $\beta_{5{\rm k}pot}=0.5{\rm k}$,$\gamma=200$mV in the experiment and $C=1.2,L=3.3, \alpha=0.2,\beta=0.5,\gamma=0.2$ in the simulation.

We observe the output w.f. $y(t)$ as a function of input $F(t)$ is not smooth in the set $C=1.2$ as compared with the case of the set $C=1$. 

\newpage
\section{Multiple period output of memristor and approximate measurement of $\omega$ of response}
\begin{figure}[htb]
\begin{minipage}[htb]{0.47\linewidth}
\begin{center}
\includegraphics[width=6cm,angle=0,clip]{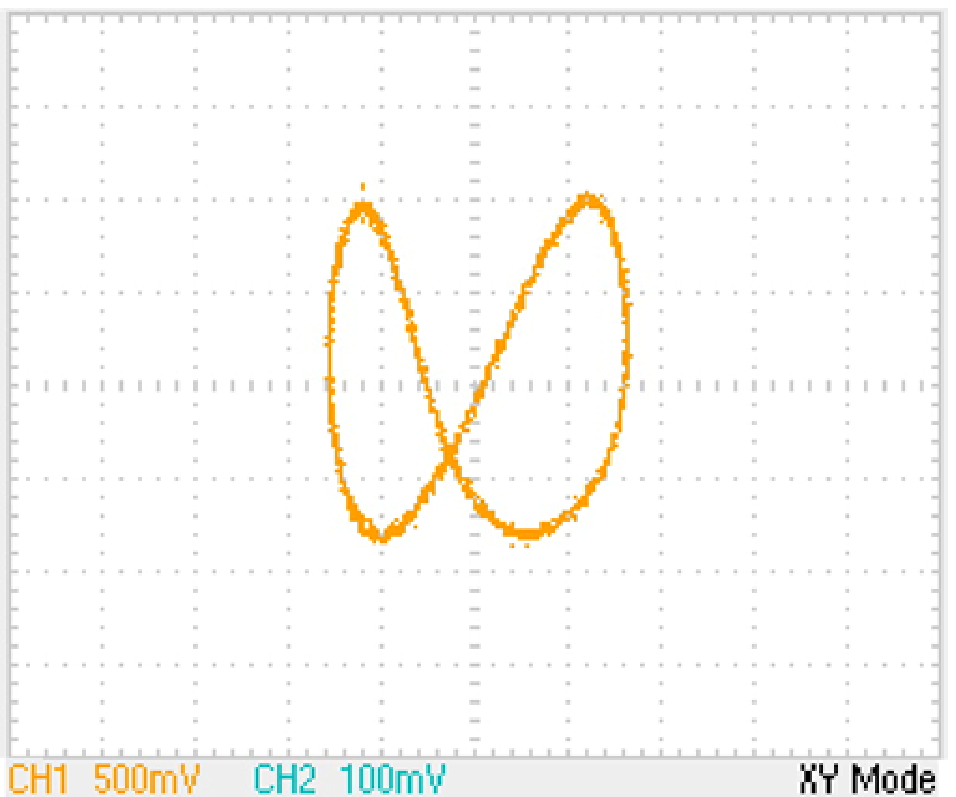}
\caption{The output w.f. $y(t)$ v.s. input w.f. $F(t)$.  $f=18.12$kHz. $C=1$}
\label{yFV3}
\end{center}
\end{minipage}
\hfill
\begin{minipage}[htb]{0.47\linewidth}
\begin{center}
\includegraphics[width=6cm,angle=0,clip]{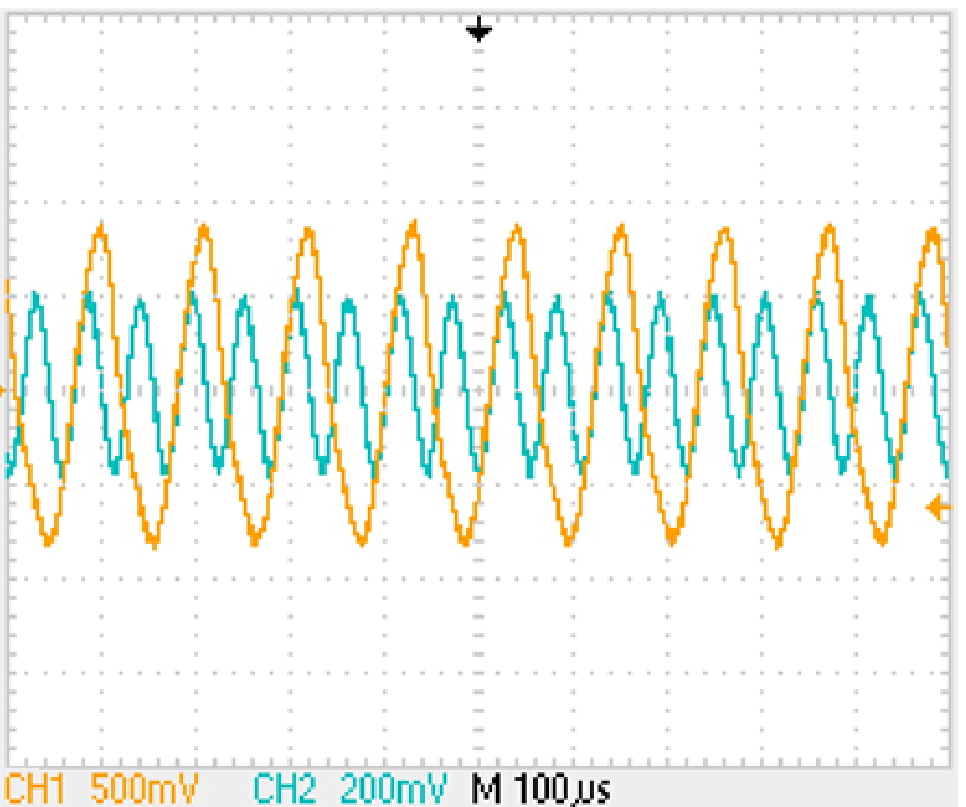}
\caption{The output w.f. $y(t)$ and input w.f. $F(t)$(green). $f=18.12$kHz. $C=1$.}
\label{yFt3}
\end{center}
\end{minipage}
\begin{center}
\includegraphics[width=12cm,angle=0,clip]{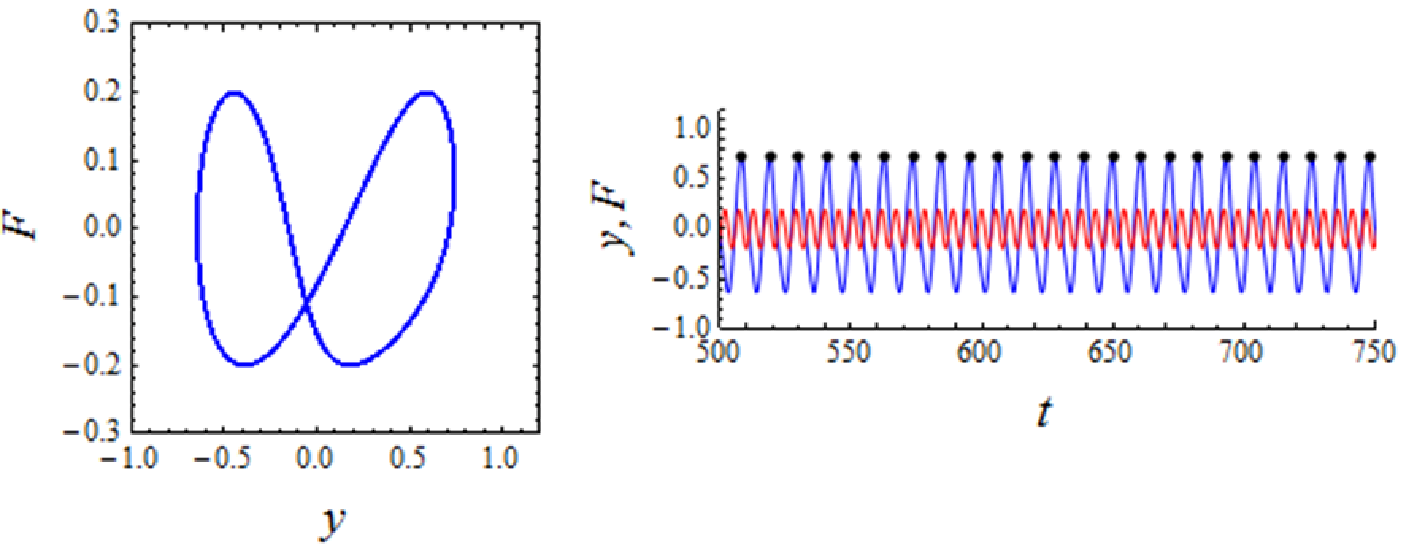}
\caption{The output w.f. $y(t)$ and input w.f. $F(t)$ (red), $C=1$.}
\label{yFsim3}
\end{center}
\end{figure}

The bifurcation diagram shows a single period output at around $\omega=1.1$rad/s. In the case of $C=1$,
the experimental results are shown in Figs.\ref{yFV3} and \ref{yFt3}. A difference of the set of these Figs.  from the set of Figs.\ref{yFV1} and \ref{yFt1} is that the input is double frequency. We assign $\displaystyle{W=\{\frac{f_s}{f_d}\}=\{\frac {2}{1}\}}$ in the present output and $\{\displaystyle\frac{1}{1}\}$ in the output of $\omega=0.59$rad/s.

The frequency $\omega$ is defined by the output number of frequency $N_d(t)$ as\cite{PRK01}
\[
\omega=2\pi f=\lim_{t\to\infty}2\pi\frac{N_d(t)}{t},
\]
and there is a proposal to measure it using Hilbert-transform\cite{Huang05}, but we measure it using a more convenient method as follows. 

We choose a relatively long period and plot the output function $y(t)$
and the input function $F(t)=\gamma\sin\omega t$. We choose two neighbouring points on which the peaks
of the two wave functions overlap, and measure this period $\tau$. We evaluate the number of frequency 
$N_d(\tau)$ that the wave $y(t)$ takes its maximum,  we make an average  $T_{d}=\tau/N_d(\tau)$ and
define  $\omega_{res}=2\pi/T_{d}$, and take it as an approximation of $\omega$. 
The numerical simulation result of $f_d=1, f_s=2$ oscillation at $C=1$ and $\omega=1.155$rad/s is given in 
Fig.\ref{yFsim3}.

We analyzed the same type of oscillation in the case of $C=1.2$ at $\omega=1.049$rad/s.
\begin{figure}[htb]
\begin{minipage}[htb]{0.47\linewidth}
\begin{center}
\includegraphics[width=6cm,angle=0,clip]{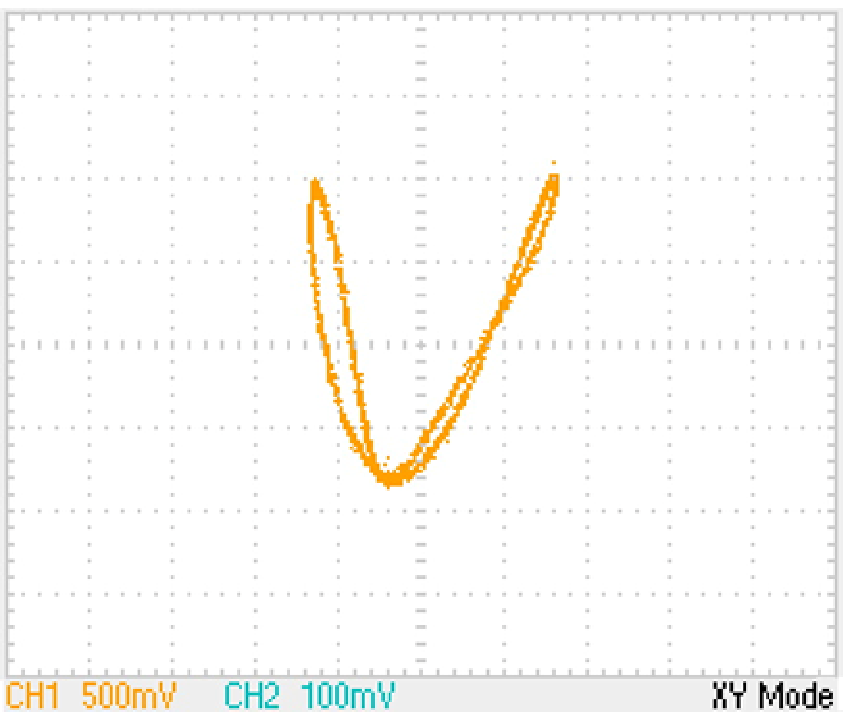}
\caption{The output w.f. $y(t)$v.s. input w.f. $F(t)$. $f=12.45$kHz. $C=1.2$.}
\label{yFV4a}
\end{center}
\end{minipage}
\hfill
\begin{minipage}[htb]{0.47\linewidth}
\begin{center}
\includegraphics[width=6cm,angle=0,clip]{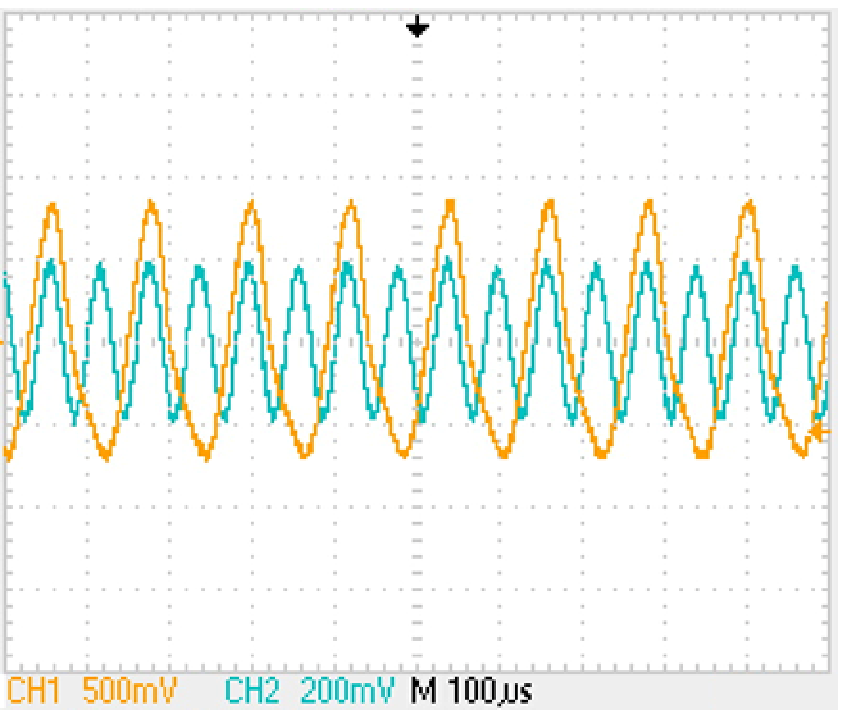}
\caption{The output w.f. $y(t)$ and input w.f. $F(t)$ (green). $f=12.45$kHz. $C=1.2$.}
\label{yFt4a}
\end{center}
\end{minipage}
\begin{center}
\includegraphics[width=12cm,angle=0,clip]{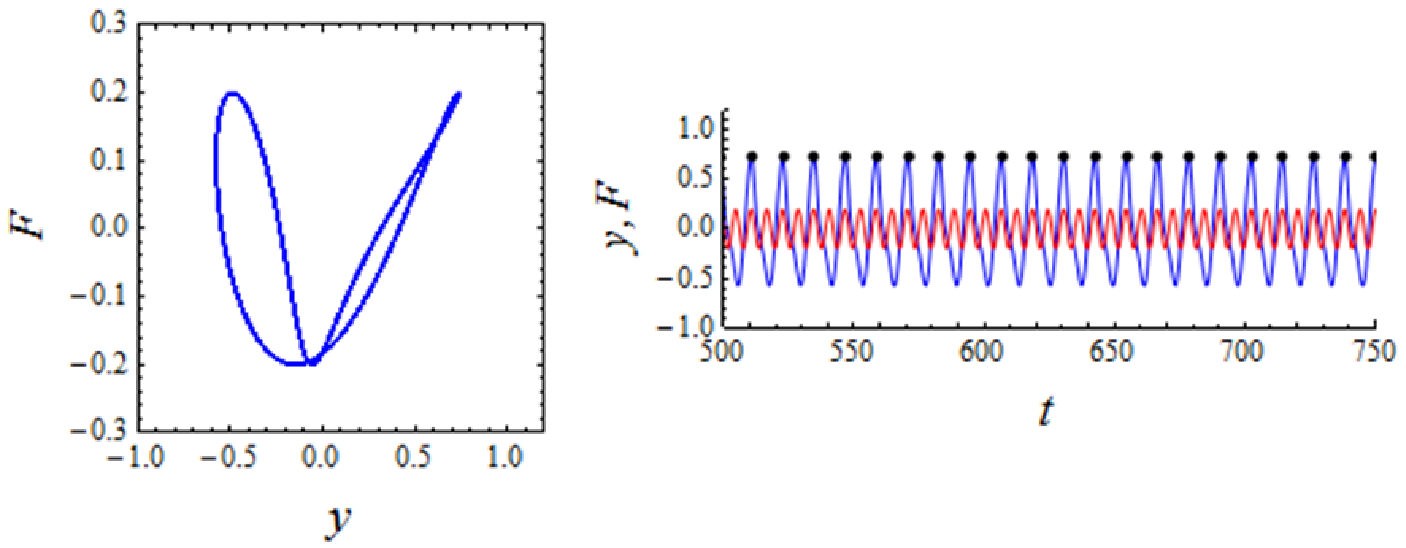}
\caption{The output w.f. $y(t)$ and input w.f. $F(t)$ (red), $C=1.2$.}
\label{yFsim4a}
\end{center}
\end{figure}

The bifurcation diagram of $C=1$ shows that there is a double period output near $\omega=0.8$rad/s. This state corresponds to the Farey sum $\displaystyle \frac{1}{1}+\frac{2}{1}\to \frac{3}{2}$, or in the Farey sequence $\displaystyle\{\frac{1}{1},\frac{3}{2},\frac{2}{1}\}$ the oscillation of $f_d=2, f_s=3$ appear between oscillations of $f_d=1, f_s=1$ and $f_d=1, f_s=2$.

The experimental results of $C=1$ at $\omega=0.86$rad/s is given in Figs.\ref{yFV4a} and \ref{yFt4a}.
The bifurcation diagrams of $C=1$ show that there is a double period output near $\omega=0.8$rad/s (cf. Fig.\ref{yFsim4a}). This state corresponds to the Farey sum $\displaystyle \frac{1}{1}+\frac{2}{1}\to \frac{3}{2}$, or in the Farey sequence $\displaystyle\{\frac{1}{1},\frac{3}{2},\frac{2}{1}\}$ the oscillation of $f_d=2, f_s=3$ appear between oscillations of $f_d=1, f_s=1$ and $f_d=1, f_s=2$.

\begin{figure}[htb]
\begin{minipage}[htb]{0.47\linewidth}
\begin{center}
\includegraphics[width=6cm,angle=0,clip]{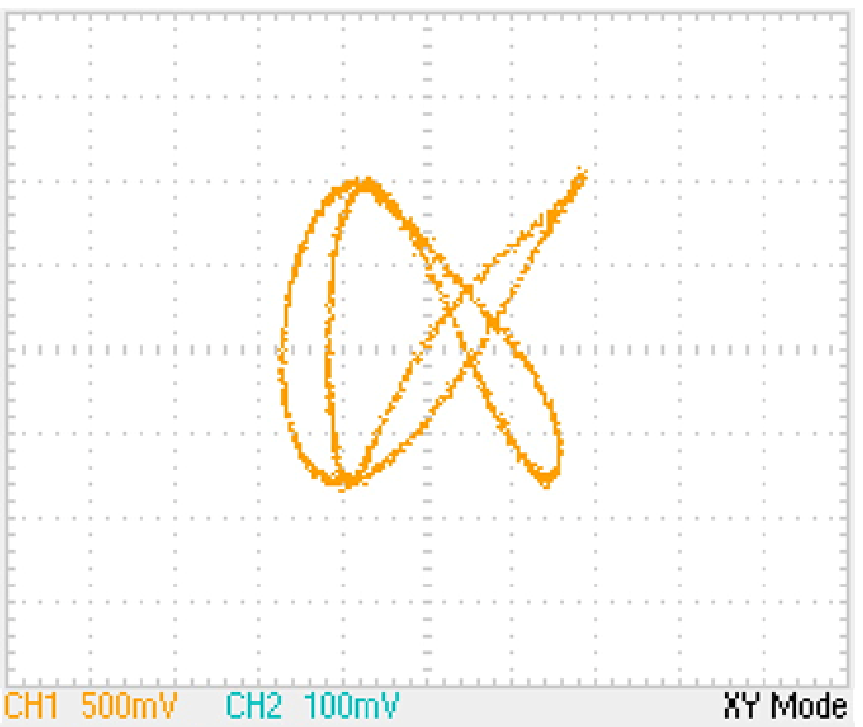}
\caption{The output w.f. $y(t)$v.s. input w.f. $F(t)$. $f=13.53$kHz. $C=1$.}
\label{yFV4}
\end{center}
\end{minipage}
\hfill
\begin{minipage}[htb]{0.47\linewidth}
\begin{center}
\includegraphics[width=6cm,angle=0,clip]{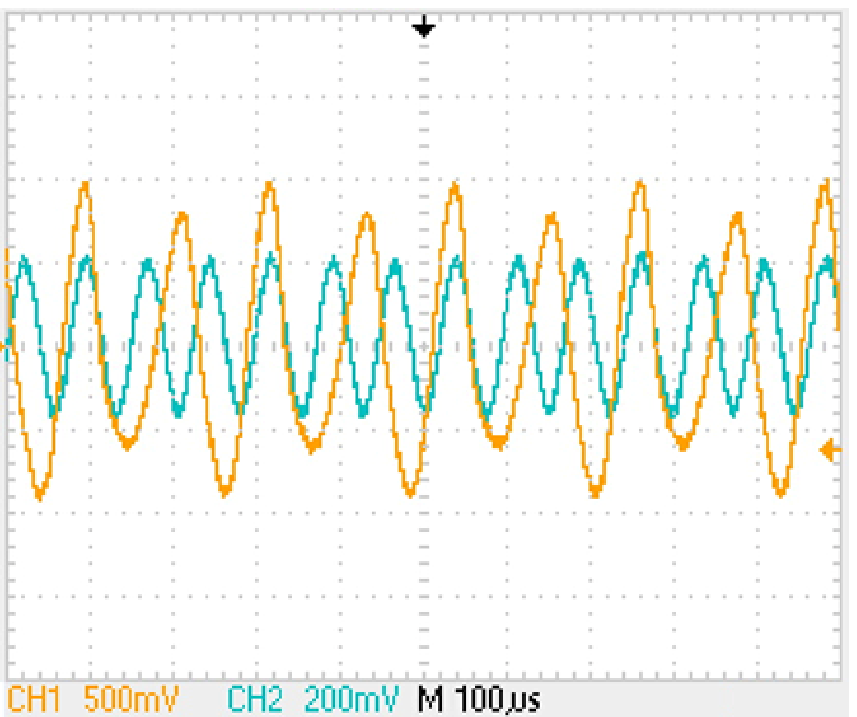}
\caption{The output w.f. $y(t)$ and input w.f. $F(t)$ (green). $f=13.53$kHz. $C=1$..}
\label{yFt4}
\end{center}
\end{minipage}
\begin{center}
\includegraphics[width=12cm,angle=0,clip]{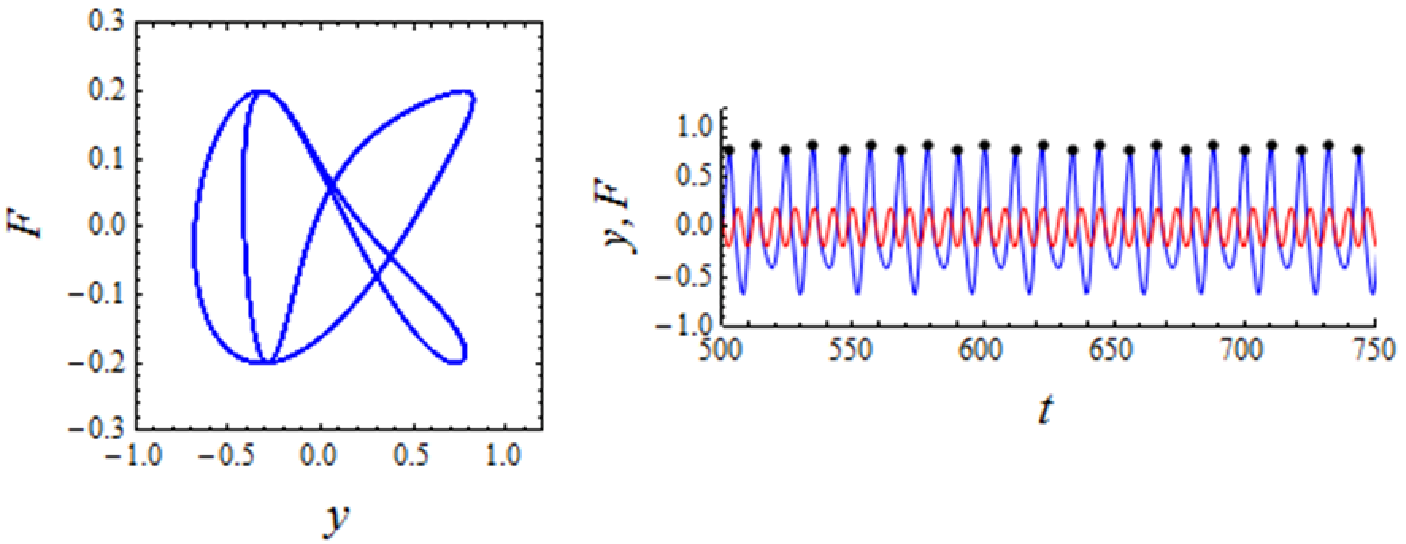}
\caption{The output w.f. $y(t)$ and input w.f. $F(t)$ (red), $C=1$.}
\label{yFsim4}
\end{center}
\end{figure}

The experimental results of $C=1$ at $\omega=0.86$rad/s are given in Figs.\ref{yFV4}, \ref{yFt4}. 
Since we have different frequencies in $F(t)$ and $y(t)$, the overlap of $F(t)$ and $y(t)$ is relatively complicated but the ratio $\displaystyle\frac{4}{3}$ can be easily checked and we can measure $\omega_{res}$ using the simulation data of Fig.\ref{yFsim4}.

\begin{figure}[htb]
\begin{minipage}[htb]{0.47\linewidth}
\begin{center}
\includegraphics[width=6cm,angle=0,clip]{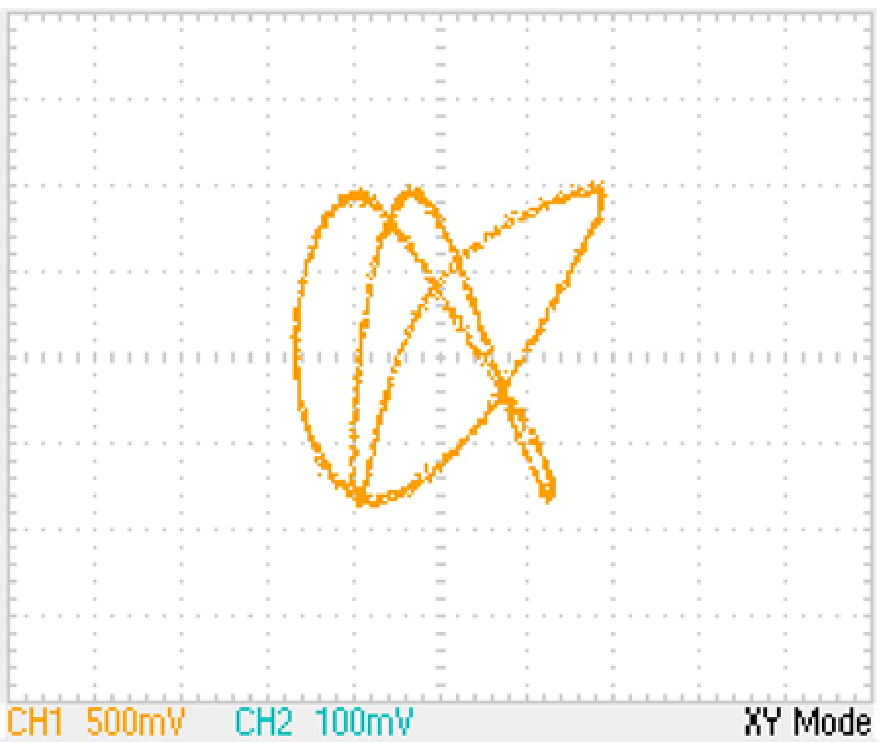}
\caption{The output w.f. $y(t)$v.s. input w.f. $F(t)$. $f=16.6$kHz. $C=1.2$.}
\label{yFV3a}
\end{center}
\end{minipage}
\hfill
\begin{minipage}[htb]{0.47\linewidth}
\begin{center}
\includegraphics[width=6cm,angle=0,clip]{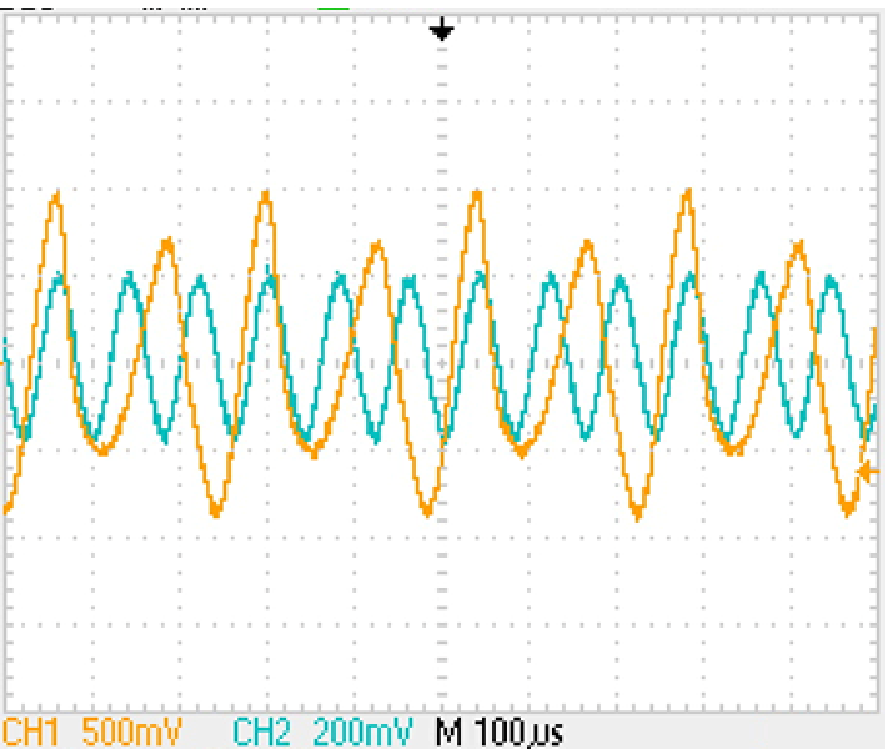}
\caption{The output w.f. $y(t)$ and input w.f. $F(t)$ (green). $f=16.6$kHz. $C=1.2$}
\label{yFt3a}
\end{center}
\end{minipage}
\begin{center}
\includegraphics[width=12cm,angle=0,clip]{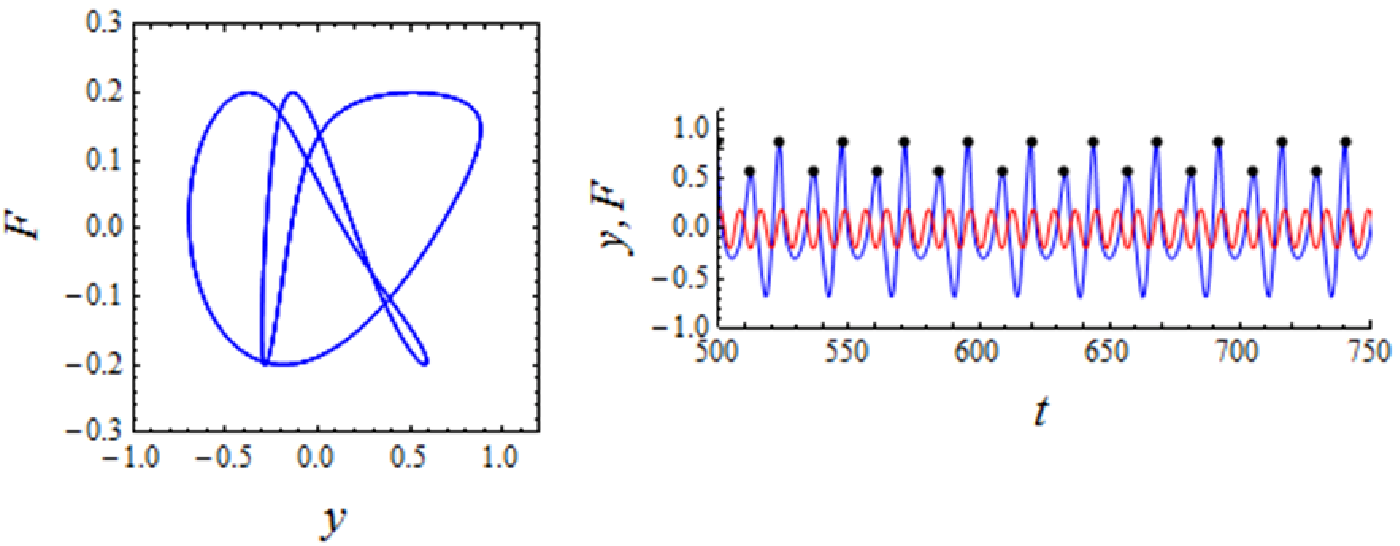}
\caption{The output w.f $y(t)$ and input w.f. $F(t)$ (red), $C=1.2$.}
\label{yFsim3a}
\end{center}
\end{figure}

We analyzed the same type of oscillation in the case of $C=1.2$ at $\omega=1.049$rad/s. 
Data are given in Figs.\ref{yFV3a} and \ref{yFt3a}.  Simulation data are Fig.\ref{yFsim3a}.

\begin{figure}[htb]
\begin{minipage}[htb]{0.47\linewidth}
\begin{center}
\includegraphics[width=6cm,angle=0,clip]{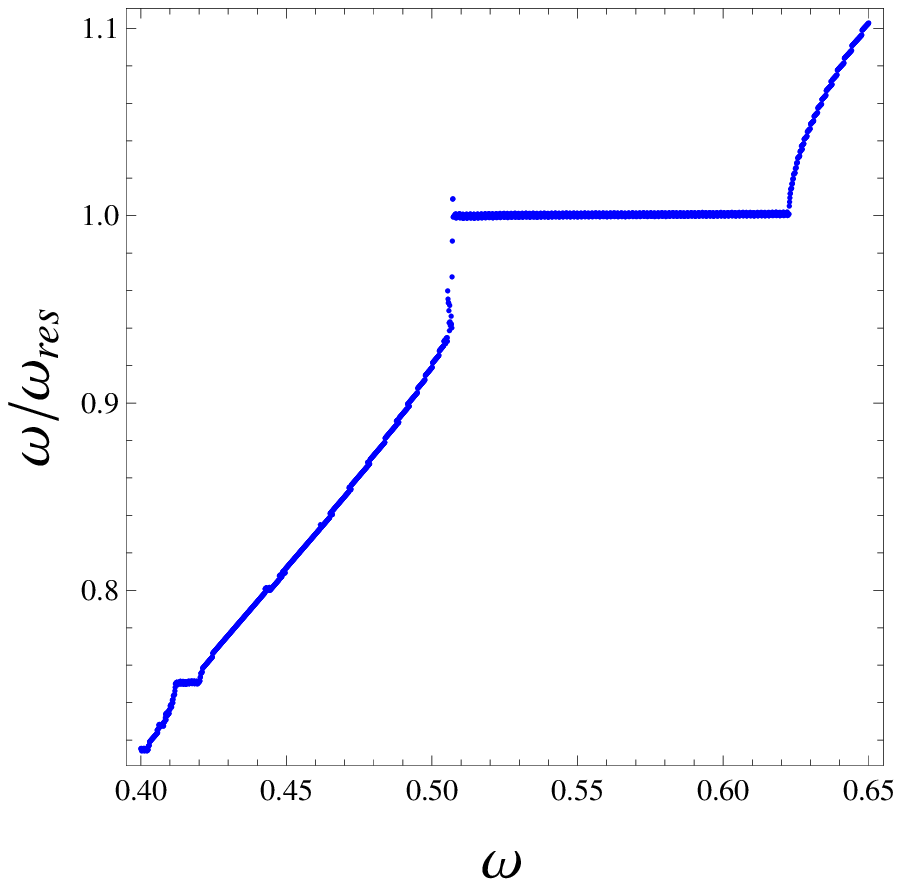}
\caption{The devil's staircase for $C=1$, where hidden attractors are observed.} 
\label{d1}
\end{center}
\end{minipage}
\hfill
\begin{minipage}[htb]{0.47\linewidth}
\begin{center}
\includegraphics[width=6cm,angle=0,clip]{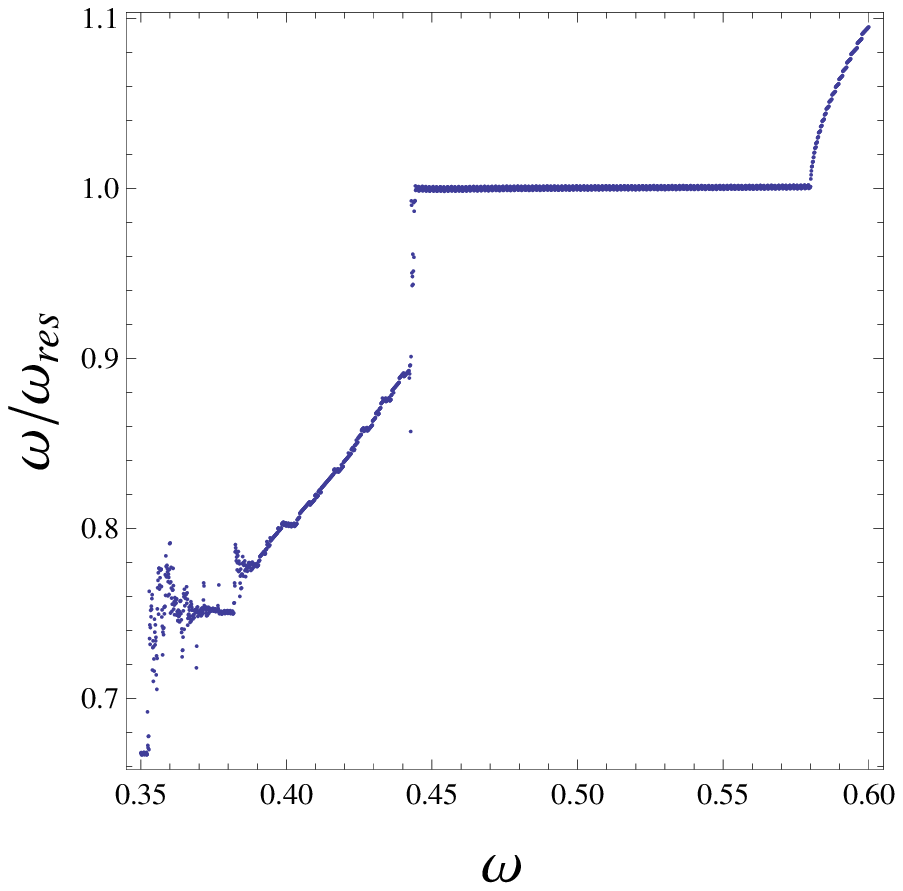}
\caption{The devil's staircase for $C=1.2$, where in $\omega<\omega_{res}$ no stable hidden attractors are found.}
\label{d2}
\end{center}
\end{minipage}
\end{figure}

\newpage
The Devil's staircases in high frequency region which have $W=3/2$ and $2/1$ of  $C=1$ and that of $C=1.2$ are qualitatively the same. Below $\omega=0.5$rad/s, in the case of $C=1$, we observe hidden attractors near $\omega=0.42$rad/s (Fig.\ref{d1}), but in the case of $C=1.2$, we do not find stable hidden attractors ( Fig.\ref{d2}).

\newpage
\section{H\"older exponent $\alpha_H$ of the difference of the rotation number $\nu$ of memristor}
When $C=1.2$, it appears Farey sequences, and it allows to measure
rotation numbers.
In the case of the model of Dos Santos, rotation number is defined by
\[
\nu(n)=\sum_{k=0}^{n-1}\frac{\Phi_{k+1}-\Phi_k}{2\pi n}.
\]

Dos Santos\cite{Santos98} and Planat\cite{Planat93} considered
\[
||\nu(x)-\nu(y)||=||x-y||^{\alpha_H},
\]
where $x$ and $y$ are two points in the space of $(c,\Omega)$. It was shown \cite{Planat93} 
that the system is stable when the H\"older exponent $\alpha_H>1$, and unstable when $\alpha_H<1$ and chaotic when $c>1$.

In the low-frequency region covered in Fig.23 of \cite{FuTa13}, i.e. the $C=1.2$ case, we take $\displaystyle x=\Omega(x)=\frac{2}{3}$ and $\displaystyle y=\Omega(y)=\frac{3}{4}$.
We define $\nu(x), \nu(y)$ as the Farey sequence
\[
 \displaystyle\{\frac{2}{3},\frac{3}{5},\frac{1}{2}\} {\quad\rm and\quad } \{\frac{3}{4},\frac{7}{9},\frac{4}{5}\},
\]
 or
\[
\displaystyle \nu(x)=\frac{2}{3}+\frac{1}{2}\to\frac{3}{5} {\quad\rm and\quad }  \nu(y)=\frac{3}{4}+\frac{4}{5}\to\frac{7}{9}.
\]

We find $\displaystyle ||\nu(x)-\nu(y)||=||\frac{7}{9}-\frac{3}{5}||=\frac{8}{45}$ and $\displaystyle ||\Omega(x)-\Omega(y)||=||\frac{2}{3}-\frac{3}{4}||=\frac{1}{12}$ yields $\displaystyle \frac{8}{45}=(\frac{1}{12})^{0.695}$ which means that the system is unstable.

Another series can be derived from the Farey sequence
\[
\displaystyle \{\frac{2}{3},\frac{5}{8},\frac{3}{5}\} {\quad\rm and\quad }  \{\frac{3}{4},\frac{7}{9},\frac{4}{5}\},
\] 
or
\[
\displaystyle \nu(x)=\frac{2}{3}+\frac{3}{5}\to\frac{5}{8} {\quad\rm and\quad } \nu(y)=\frac{3}{4}+\frac{4}{5}\to\frac{7}{9}.
\]
In this case
$\displaystyle ||\nu(x)-\nu(y)||=||\frac{7}{9}-\frac{5}{8}||=\frac{11}{72}$ yields  $\displaystyle\frac{11}{72}=(\frac{1}{12})^{0.756}$ 
which means that the system is unstable too.

In the low-frequency region covered in Fig.22 of \cite{FuTa13}, i.e. of $C=1$ case, we have
\[
 \displaystyle\{\frac{2}{3},\frac{5}{7},\frac{3}{4}\} {\quad\rm and\quad } \{\frac{1}{2},\frac{3}{5},\frac{2}{3}\},
\]
 or
\[
\displaystyle \nu(x)=\frac{2}{3}+\frac{3}{4}\to\frac{5}{7} {\quad\rm and\quad }  \nu(y)=\frac{1}{2}+\frac{2}{3}\to\frac{3}{5}.
\]
The level $\displaystyle\frac{5}{7}$ is not clear in \cite{FuTa13}, but a fine bifurcation diagram shows that it exists between levels of $\displaystyle\frac{2}{3}$ and $\displaystyle\frac{3}{4}$.
In this case
$\displaystyle ||\nu(x)-\nu(y)||=||\frac{5}{7}-\frac{3}{5}||=\frac{4}{35}$ and  $\displaystyle ||\Omega(x)-\Omega(y)||=||\frac{2}{3}-\frac{1}{2}||=\frac{1}{6}$  yields  $\displaystyle\frac{4}{35}=(\frac{1}{6})^{1.211}$,
which means that the system is stable.

This difference of $\alpha_H$ explains that the system of $C=1.2$ is chaotic, but $C=1$ is not.

\section{Discussion and Conclusion}
We studied the devil's staircase structure in the Muthuswamy-Chua's 
memristic circuit perturbed by a sinusoidal circuit.  When the amplitude of the external current is $C=1.2$,
we observed chaotic behavior superposed on the devil's staircase. Nevertheless, when it is $C=1$, standard devil's staircase structure is observed. 

In the case of Dos Santos's model, $\Phi$ changed in the region [0, $2\pi$] and the number of solution of $1-c\sin\Phi=0$ which depends on $c>1$ or $c<1$ was important. In the case of memristic current,
the qualitative difference of the rotation number $\nu(x)$ as the amplitude of external perturbation 
occurs from qualitative difference of the time dependence of current $y(t)$ in the case of $C=1.2$ and $C=1$.
We observed Farey's sequence of $\displaystyle W=\frac{q+Q}{p+P}$ in the ratio of the frequency of the driving oscillation $f_s$ and that of the response $f_d$.  Furthermore, we measured the difference of the rotation number $||\nu(x)-\nu(y)||$ and that of $||\Omega(x)-\Omega(y)||$ and $\alpha_H=\log_{||\Omega(x)-\Omega(y)||}(||\nu(x)-\nu(y)||)$.

 In the case of $C=1$ we obtained $\alpha_H>1$, while in the case of $C=1.2$ we obtained $\alpha_H<1$. The system of $C=1.2$ is unstable and chaos is observed, while in the case of $C=1$ clear devil's staircase was observed.  
The origin of this difference is not clear, but it coincides with the difference of the structure of output w.f. $y(t)$ as a function of input w.f. $F(t)$.

\vskip 0.5 true cm
\leftline{\bf Acknowledgments}

\noindent We thank Dr. Serge Dos Santos for sending his PhD theses, which contains valuable information on the chaos via devil's staircase, and giving us helpful comments.


\begin{thebibliography}{9}
\bibitem{Santos98} Dos Santos, S. , "\'Etude non lin\'eaire et arithm\'etique de la synchronisation des syst\`emes: application aux fluctuations de basse fr\'equence des oscillateurs ultra-stables", {\it Th\`ese Grade de Docteur de l'Universit\'e de Franche-Comt\'e } (1998). 
\bibitem{Devaney89} Devaney, R.L., "An Introduction to Chaotic Dynamical System", Perseus Books Publishing, L.L.C.,  Sect.1.15 (1989) .
\bibitem{CYY86} Chua, L.O.,  Yao, Y. and Yang, Q. "Devil's Staircase Route to Chaos in a Non-Linear Circuit", Circuit Theory and Applications, {\bf 14}, pp.315--329, (1986).
\bibitem{MuCh10} Muthuswamy, B and  Chua, L.O., "Simplest Chaotic Circuit", International Journal of Bifurcation and Chaos, {\bf 20}, pp.1567--1580 (2010).
\bibitem{ChKa76} Chua,L.O. and Kang, S.M. ,"Memristive Devices and Systems", Proceedings of the IEEE, {\bf 64}, pp.209--223,(1976).
\bibitem{GLC10} Ginoux,J.M., Letellier, Ch.  and Chua, L.O. , "Topological Analysis of Chaotic Solution of Three-Element Memristive Circuit",
International Journal of Bifurcation and Chaos {\bf 20}, pp.3819--3827, (2010).
\bibitem{LlVa12} Llibre,J. and Valls,C., "On the Integrability of a Muthuswamy-Chua System", Journal of Nonlinear Mathematical Physics, {\bf 19},pp.1250029-1250041, (2012).
\bibitem{ZhZh13} Zhang, Y and Zhang, X. ,"Dynamics of the Muthuswamy-Chua System", International Journal of Bifurcation and Chaos, {\bf 23}, pp. 1350136-1350143 (2013). 
\bibitem{FuTa13} Furui,S. and Takano,T., "Strange Attractors of Memristor and Devil's Staircase Route to Chaos", arXiv:1312.3001, (2013). 
\bibitem{PRK01} Pikovsky,A., Rosenblum,M. and Kurths,J. ,
"Synchronization -A Universal Concept In Nonlinear Sciences-", Cambridge Unversity Press, Sect.10.2,(2001).
\bibitem{Huang05} Huang, N.E. , "Hilbert-Huang transform", http://www.schlarpedia.org/article/Hilbert-Huang transform (2005).
\bibitem{Planat93} Planat, M. and Koch, P. , "Plurifractal Signature in the Study of Resonances of Dynamical Systems", Fractals, vol.1 pp.727--734, (1993).

\end{thebibliography}
\end{document}